\documentclass[sigconf]{acmart}
\usepackage{multirow} 
\usepackage{float}
\usepackage{subfig}                 %
\usepackage{makecell}
\usepackage{algorithm}
\usepackage{algpseudocode}  
\usepackage{amsmath}
\usepackage{colortbl}

\AtBeginDocument{%
  }

\setcopyright{acmlicensed}
\copyrightyear{2018}
\acmYear{2018}
\acmDOI{XXXXXXX.XXXXXXX}

\acmConference[CIKM'24]{}{}{}

\renewcommand\footnotetextcopyrightpermission[1]{}

\newcommand{\method}{\textsc{ECLIPSE}}
\newcommand{\bench}{\textsc{ECLIPSE-Bench}}

\begin{document}


\title{ECLIPSE: Semantic Entropy-LCS for Cross-Lingual Industrial Log Parsing}

\author{Wei Zhang\textsuperscript{\textrm{1}}, Xianfu Cheng\textsuperscript{\textrm{1}}, Yi Zhang\textsuperscript{\textrm{1}}, Jian Yang\textsuperscript{\textrm{1}†},  Hongcheng Guo\textsuperscript{\textrm{1}}, Zhoujun Li\textsuperscript{\textrm{1}†}, \\ Xiaolin Yin\textsuperscript{\textrm{2}}, Xiangyuan Guan\textsuperscript{\textrm{3}}, Xu Shi\textsuperscript{\textrm{3}}, Liangfan Zheng\textsuperscript{\textrm{3}}, Bo Zhang\textsuperscript{\textrm{3}}}
\affiliation{
  \institution{\textsuperscript{\textrm{1}}State Key Laboratory of Complex \& Critical Software Environment, Beihang University}
  \country{}
}
\affiliation{
  \institution{\textsuperscript{\textrm{2}}Haier Smart Home, \textsuperscript{\textrm{3}}Cloudwise Research}
  \country{}
}
\email{{zwpride, buaacxf, zhangyi2021, jiaya, hongchengguo, lizj}@buaa.edu.cn;}
\email{yinxiaolin@haier.com; {tim.shi, leven.zheng, bowen.zhang}@cloudwise.com;}

\renewcommand{\shortauthors}{Trovato et al.}

\begin{abstract}

Log parsing, a vital task for interpreting the vast and complex data produced within software architectures faces significant challenges in the transition from academic benchmarks to the industrial domain. Existing log parsers, while highly effective on standardized public datasets, struggle to maintain performance and efficiency when confronted with the sheer scale and diversity of real-world industrial logs. These challenges are two-fold: 1) massive log templates: The performance and efficiency of most existing parsers will be significantly reduced when logs of growing quantities and different lengths; 2) Complex and changeable semantics: Traditional template-matching algorithms cannot accurately match the log templates of complicated industrial logs because they cannot utilize cross-language logs with similar semantics. 
To address these issues, we propose \textbf{\method{}}, \textbf{E}nhanced \textbf{C}ross-\textbf{L}ingual \textbf{I}ndustrial log \textbf{P}arsing with \textbf{S}emantic \textbf{E}ntropy-LCS, since cross-language logs can robustly parse industrial logs. On the one hand, it integrates two efficient data-driven template-matching algorithms and Faiss indexing. On the other hand, driven by the powerful semantic understanding ability of the Large Language Model (LLM), the semantics of log keywords were accurately extracted, and the retrieval space was effectively reduced. Notably, we launch a Chinese and English cross-platform industrial log parsing benchmark \bench{} to evaluate the performance of mainstream parsers in industrial scenarios. Our experimental results across public benchmarks and \bench{} underscore the superior performance and robustness of our proposed \method{}. Notably, \method{} both delivers state-of-the-art performance when compared to strong baselines and preserves a significant edge in processing efficiency\footnote{We will release our code and dataset.}.

\end{abstract}



\maketitle
{
 \let\oldthefootnote\thefootnote
  \let\thefootnote\relax\footnotetext{† Corresponding author.}
  \let\thefootnote\oldthefootnote
}

\section{Introduction}
\begin{figure}[tp]
\centering
\includegraphics[width=8cm]{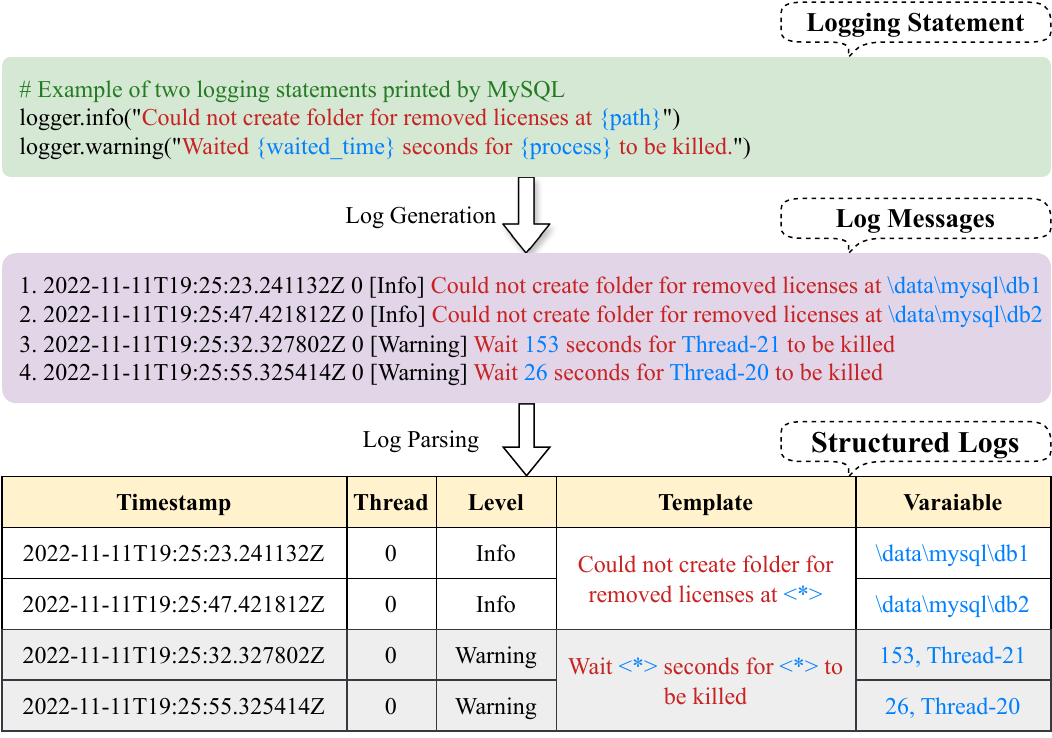}
\caption{Diagram of the log structuring process. The semi-structured logs are generated by log statement code and then parsed into the structured logs by algorithms. The log template represents the part of the keywords that are not changed during the log parsing process.}
\label{fig1}
\vspace{-15pt}
\end{figure}

Logs play a crucial role in Algorithmic IT Operations (AIOps) behavior compared to other operational and maintenance data, as they provide significant insights into system behavior \cite{automated_log_analysis,tools_and_benchmarks}. By analyzing logs, we can complete a variety of downstream tasks, such as anomaly detection \cite{deeplog,cfg_mining,zhang2024lemur}, fault diagnosis \cite{fault_identifying,zhang2024mabc} and root cause analysis \cite{root_logan,root_diagnosing}. In general, most log analysis methods make log parsing the primary step in automated log analysis \cite{he2016evaluation}. Illustrated in~\ref{fig1}, semi-structured logs are generated from log statement code and need to be parsed to structured logs by algorithms. The most challenging task among them is to extract templates from logs \cite{tools_and_benchmarks}, which represent the unchanged parts of the log parsing process. Traditional log parsing was accomplished by manually configuring regular expressions \cite{using_regex}. With the exponential growth in the volume and complexity of logs, cloud computing has rendered this approach impractical. Another method is source code-based log parsing \cite{using_source_code1, using_source_code2}, which are widely used as third-party libraries and some library source codes may not be accessible.

    
Many data-driven log parsing techniques have been proposed. Some are based on the assumption that constants appear more frequently than variables in log messages \cite{slct,logcluster,logram}, while others treat log parsing as a clustering problem \cite{logmine, shiso, lenma, lpv} or use specific heuristic rules \cite{iplom, spell, drain,prefix_graph}. Although these methods theoretically make log parsing feasible for industrial applications, most have only been tested on simple public datasets and lack practical experience with complex industrial logs. Therefore, they actually perform poorly in most industrial logs, with both efficiency and accuracy being low. Our paper addresses this gap by identifying the following issues overlooked by current log parsing algorithms based on an analysis of industrial logs and existing algorithms: 1) \textbf{Enormous number of log templates:} A massive number of log templates on parsing efficiency, especially in industrial applications where heuristic rules used to accelerate computation cannot fundamentally solve efficiency problems. For example, Spell \cite{spell} and Drain\cite{drain} perform poorly in industrial log template scenarios due to frequent computation. 2) \textbf{Length-Insensitivity:} Existing Log parsing algorithm assumes that the prior is generally the same keyword, which makes it insensitive to length. Drain+\cite{drain+} was aware of this issue and raised Jaccard similarity to merge logs with the same token, disregarding the order and position of identical tokens. 3) \textbf{Complex semantics and multilingual logs:} In industrial scenarios, logs from different programming languages, system sources, and language types often aggregate under the same data source. Data-driven log parsing methods are capable of handling the former, but cannot handle the latter.

To address these issues, we propose a powerful log parser called \method{}. Specifically, driven by strong keyword representations, cross-linguistic understanding, and cross-semantic comprehension of the large language model (LLM), \method{} builds a dynamic dictionary from semantic keywords to log templates. K-nearest neighbor templates will be recalled from templates in the dynamic dictionary by Faiss. Then, the carefully designed method Entropy-LCS, an entropy-improved longest command subsequence, will identify candidates as log templates for updating the dynamic dictionary in real time. 
To construct a benchmark called \bench{}, we align with industrial applications compared with public benchmark datasets to further validate the effectiveness of our algorithm. The experiment results demonstrate its effectiveness on both public benchmarks and the \bench{} benchmark. Besides, we also explore the parsing efficiency comparison between \method{} and other strong baseline algorithms, where there are a large number of templates involved.
    
Overall, our main contributions can be summarized as follows,

    \begin{itemize}
    \item We propose a powerful log parsing system called \method{}. \method{} creatively employs LLM to drive cross-linguistic and cross-semantic relevance detection and ensures efficient retrieval of logs under massive template influence by Faiss addressing the challenges in log parsing.

    \item Using Entropy-LCS, an information entropy-improved LCS, for real-time matching of log intrinsic K-nearest neighbor templates not only effectively solves the problem of length insensitivity, but also further improves the performance and flexibility of log parsing.
    
    \item We construct \bench{}, a bilingual industrial log parsing benchmark dataset, which supports both Chinese and English languages. We collected nearly 102M logs from 3 industrial domains and extracted 700 templates from these logs. We evaluate \method{} on both public Loghub and \bench{} using F-measure, grouping accuracy, and execution time, and \method{} achieve advanced performance and highly competitive parsing efficiency. Specifically, it outperforms the current existing industry-standard baseline methods.

    \end{itemize}

\section{Related Work}

\subsection{Existing matching strategies}
A lot of work has been proposed in the field of log parsing in recent years. In general, log parsing can be classified into four categories: frequent pattern mining, clustering, heuristics rules, and deep learning methods. \textbf{1) Frequent pattern mining:} This category assumes that constants generally occur more frequently in log messages than variables. SLCT \cite{slct} and LogCluster \cite{logcluster} mainly group logs into several clusters based on the frequency term of each log. Logram \cite{logram} uses frequent n-grams to separate constants and variables. \textbf{2) Clustering:} This category usually clustering logs, then extracts templates from each cluster. LogMine \cite{logmine} and SHISO \cite{shiso} use hierarchical clustering to cluster logs and update the template. LKE \cite{lke} employs edit distance and k-means to cluster logs. LenMa \cite{lenma} helps cluster logs based on length vector. LPV \cite{lpv} uses the semantic vectors generated by Word2vec to cluster the logs. \textbf{3) Heuristics rules:} This category uses some heuristic rules to help log parsing. For example, IPLOM\cite{iplom} iteratively classifies logs according to length, token at a specific position, and mapping relation. Spell\cite{spell} regards log parsing as the LCS problem. Drain\cite{drain} utilizes length and prefix tokens to partition logs. Recently, Preﬁx-Graph\cite{prefix_graph} proposes that generate log templates form a prefix graph. \textbf{4) Deep Learning methods:} In recent years, deep learning-based log parsing algorithms have emerged. For example, Nulog \cite{nulog} uses a Transformer-based encoder layer to train a log parsing network. Uniparser \cite{uniparser} designs three modules and uses contrastive learning for log parsing. LogAP\cite{logap} employs Machine Translation (MT) to perform parsing tasks. However, deep learning algorithms are usually inefficient and costly for training and inference. Therefore, it is still challenging to apply these methods in real scenarios.
        
\subsection{LLM in log semantic parsing} With the rapid advances in language modeling \cite{owl,transformer,softtemplate,shen2023hugginggpt,zhang2024mabc,wang2022self,bai2023qwen,yao2023react,shinn2023reflexion,xnlg}, and particularly the emergence of LLMs with Transformer-based architectures such as GPT-3.5, GPT-4 \cite{openai2023gpt4} and PaLM \cite{palm2}, Its excellent language understanding, generation, generalization, and reasoning capabilities greatly promote the integration of Natural Language Processor (NLP) and AIOps tasks. The integration of LLM with external databases and APIs further enhances its functionality\cite{xcot, owl, kojima2022large, wang2022self}, so that domain-specific knowledge can be more effectively integrated and continuously updated, especially when applied to the semantic analysis of logs, which greatly improves the accuracy of log parsing and anomaly detection \cite{code_llama, zhang2024lemur, guo2023loglg}. In addition, Retrieval Augmented Generation technology enables LLM to have the ability to access external knowledge sources, even when faced with more complex and knowledge-intensive tasks, generating answers that are more factual, specific, and diverse\cite{lewis2021retrievalaugmented,bai2023qwen,yao2023react}. In our work, Using LLM to reorder the representation vector of log keywords to highlight the most relevant results, and then with the powerful background knowledge of LLM, the most credible template matching length is selected for the result sequence, which effectively reduces the number of templates that need to be matched in the current input log and realizes the dual purpose of information retrieval enhancer and filter. Provide refined inputs for more accurate log parsing algorithms.


\section{ANALYSIS ON INDUSTRIAL LOG PARSING CHALLENGES}
In this section, we aim to provide a detailed analysis of the issues mentioned earlier and use practical examples to illustrate the causes behind them.

\subsection{Huge volume of log templates}
The quantity level of industrial logs exceeds that of public data by one order of magnitude, and the resulting number of log templates is also relatively large. Illustrated in Table~\ref{table1}, we analyzed the official website and source code of open-source software and found that the number of log templates exceeded 6000. At the same time, the reason for the small number of log templates in the public dataset Loghub is that it comes from sampling, and in non-sampling situations, its log templates will also be equally large.

\begin{table}[htbp]
\caption{Number of templates of some open source software.}
\vspace{-10pt}
\begin{center}
\resizebox{0.9\columnwidth}{!}{\begin{tabular}{cccccc}
   \toprule
   Source & MySQL&Oracle&Cisco&ClickHouse\\
   \midrule
   Number of templates & 4251&648&1556&5997\\
   \bottomrule
\end{tabular}}
\end{center}
\vspace{-10pt}
\label{table1}
\end{table}

The time complexity of a cluster-based algorithm is typically represented as $O(mn)$ or $O(mlog(n))$, where $m$ denotes the data size and $n$ denotes the number of clusters. In log parsing, the number of clusters is the number of templates. Therefore, it is essential to consider the impact of the number of log templates for log parsing.

\subsection{Various lengths of logs from the same template}

\begin{figure}[htp]
\centering
\includegraphics[width=7.95cm]{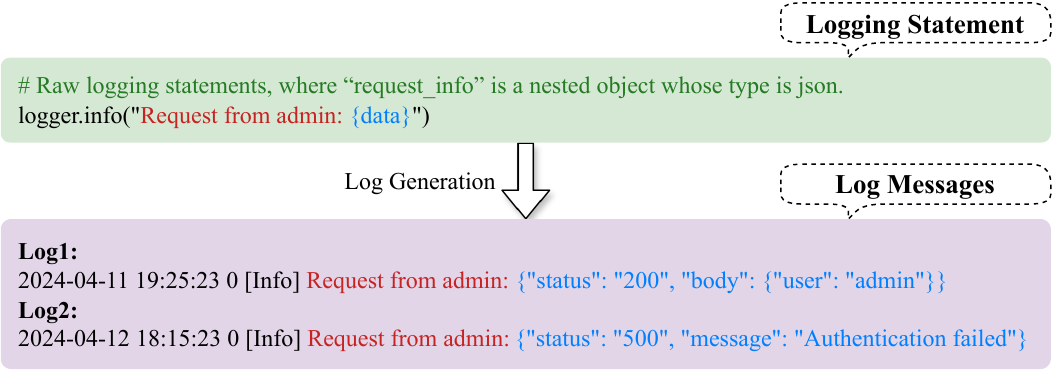}
\vspace{-5pt}
\caption{Example of various lengths of two logs from the same template.}
\label{fig3}
\vspace{-10pt}
\end{figure}

Many algorithms assume that logs belonging to the same template must have the same length, but this is not always the case in complex industrial logs, especially those with complex nested objects like JSON, XML, etc \cite{low_resource_template,alm,soft_template}. Illustrated in Figure~\ref{fig3}, $Log1$ and $Log2$ are printed by the same code, in other words, they belong to the same template. However, their lengths vary a lot due to the significant differences in the field named "BUSI\_INFO" of JSON objects. This happens very frequently, but most algorithms don't have the ability to handle this issue because they regard the same log length as a prerequisite for logs to belong to the same template, like Drain\cite{drain}, IPLOM\cite{iplom}, Lenma\cite{lenma}, and so on.

\begin{figure}[htp]
\centering
\includegraphics[width=8cm]{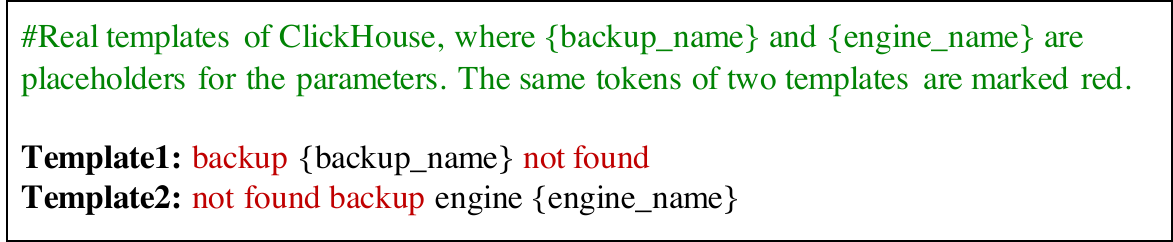}
\vspace{-5pt}
\caption{Example of templates with similar tokens but different orders.}
\label{fig4}
\vspace{-10pt}
\end{figure}

\begin{figure}[htp]
\centering
\includegraphics[width=8.5cm]{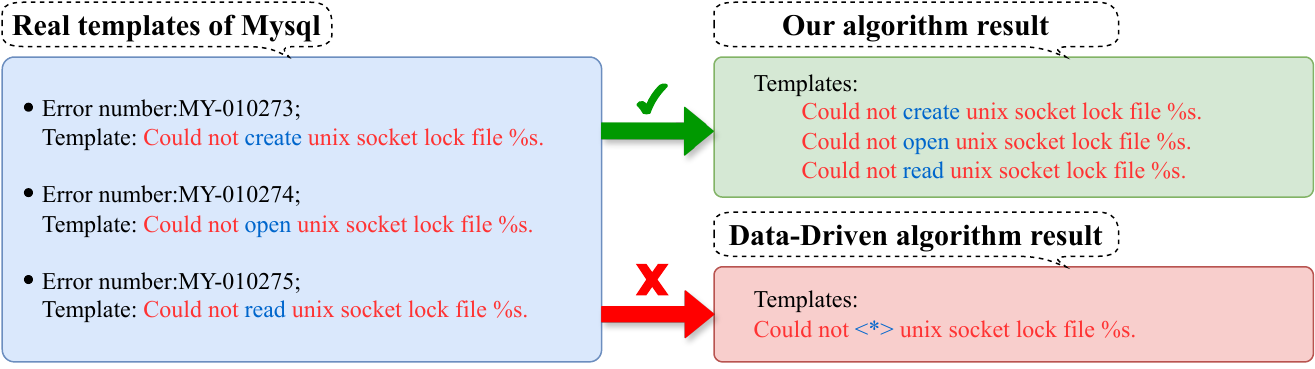}
\vspace{-5pt}
\caption{Example of three similar log templates in MySQL error log.}
\label{fig2}
\vspace{-10pt}
\end{figure}
    
We are glad to see Drain+\cite{drain+} has also realized this issue, and to address it, Drain+ first uses Drain\cite{drain} to generate a series of templates and then merges them by comparing their Jaccard Similarity. However, the flaw of this approach is that Jaccard similarity only considers the number of identical tokens between two logs but ignores the order and position of identical tokens. So templates with similar tokens but different orders of tokens may be merged by it, just as illustrated in Figure~\ref{fig4}. The two templates both have tokens "backup", "not" and "found", but their orders are totally different, which may cause a big problem in extracting parameters from the log. This type of problem appears more in more non-standard logs and more in Chinese, Japanese, Korean, or other languages with more flexible syntax than English.

\subsection{Others}
In addition to these two main issues that may significantly impact log parsing, most algorithms also often neglect the semantic differences between logs with similar characters. Logs reflect system behavior, and some events in the system are closely related. Therefore, developers often write similar-looking logs to maintain consistency. To illustrate this point, we take MySQL error logs as an example. We found three log templates that look very similar on the official website. In Figure~\ref{fig2}, these three templates describe three totally different operations performed on the Unix socket lock file. However, in the data-driven algorithm, they are likely to be considered as one template. Similar situations can also occur when describing the same operations on different objects and so on. We analyzed the templates of MySQL error logs with error numbers MY-010000 to MY-010100 and found that around one-fifth of the templates have similar counterparts, which shows this is also a widespread problem.




\begin{figure*}[htp]
\centering
\includegraphics[width=0.95\textwidth]{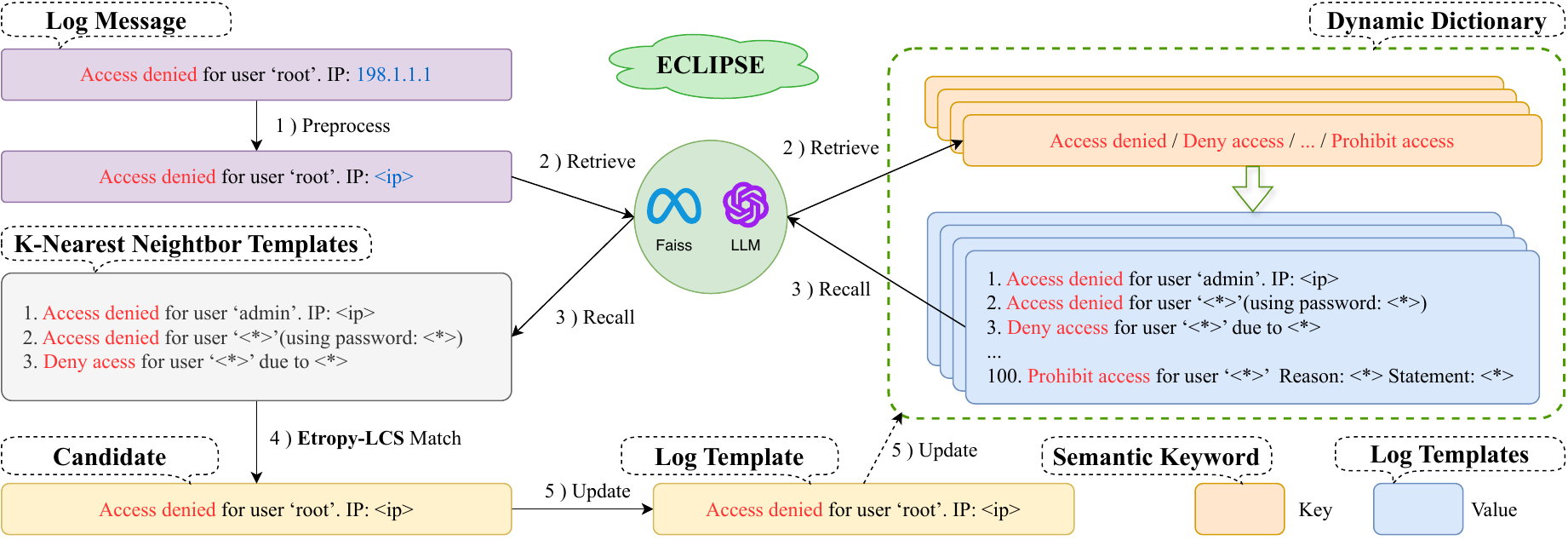}
\caption{The framework of \method{}. It is a five-stage two-path cross-lingual online log parser. Specifically, driven by LLM, \method{} constructs a dynamic dictionary from semantic keywords to log templates. Faiss recalls K-nearest neighbor templates from the dynamic dictionary. Then, the Entropy-LCS method will identify candidates as log templates and log templates will update the dynamic dictionary in real time.}
\label{fig5}
\vspace{-10pt}
\end{figure*}

\section{Methodology}
\label{method}
\method{} defines parsing as a template retrieve problem. To achieve parsing \method{} utilizes a special dynamic dictionary structure called template library to restore seen templates. As each log arrives, \method{} aims to retrieve the most suitable template and then update our template library. In this chapter, we will describe in detail how \method{} works.

\subsection{Overview of \method{}}
Fig~\ref{fig5} illustrates how \method{} uses a dynamic dictionary structure to store templates of previously seen logs. This structure takes log semantic keywords as keys and each key maps to a value containing a list of templates and a Faiss index. Our algorithm operates and starts with an empty dictionary. When a new log is coming, \method{} takes five steps to identify the most suitable template:
1) \textbf{Preprocess} the log.
2) Extract the semantic keywords from the log, and take the log semantic keywords as the key to \textbf{retrieve} log templates and a Faiss index in the dynamic dictionary.
3) Utilize the corresponding Faiss index to \textbf{recall} k-nearest-neighbor templates.
4) Using \textbf{Entropy-LCS} to \textbf{match} the log to k-nearest-neighbor templates to choose the most suitable candidate.
5) \textbf{Update} the corresponding log templates and Faiss index.

\subsection{Step 1. Preprocess}\label{Preprocess}
When a new log $l_i$ arrives, \method{} preprocesses it before template searching. Previous work has proved that preprocessing can effectively improve parsing accuracy. So in our algorithm, we also use some simple regex rules to replace some special objects with tags before template searching. Special objects include IP, URL, time, and so on. For example, in Fig \ref{fig5}, we replace "198.1.1.1" with tag "$<$ip$>$". After replacing, \method{} adopts some simple delimiters like space and colon to split logs into a token sequence $s_i$.

\subsection{Step 2. Retrieve}\label{Retrieve}
As mentioned previously, most algorithms have ignored the semantic differences between similar logs, we also gave an example in Fig~\ref{fig2} to explain. However, almost all semantic-based semantic algorithms have efficiency problems, and they may not be able to solve the problem shown in Fig~\ref{fig2}.

To balance efficiency and effect, we adopt a compromise plan. Firstly, we use LLM to process our collected logs and identify the most commonly used and important words and expressions through a keyword extraction mission. In Figure~\ref{fig5}, when the keyword library contains  "access", "denied" "open" and "close", \method{} will extract the keywords "Access denied" from the log "Access denied for user ‘root’. IP: $<$ip$>$" as its log abstract. Then, we construct a special dynamic dictionary by using semantic keywords as keys and the corresponding log templates as values.

    


\subsection{Step 3. Recall}\label{Recall}


Faiss is an open-source tool to retrieve the most similar neighbor vector efficiently from Meta, which is competent in handling efficiency issues for a huge volume of templates in industrial logs. We need to embed logs into vectors first.

\begin{figure}[htp]
\vspace{-10pt}
\centering
\includegraphics[width=7cm]{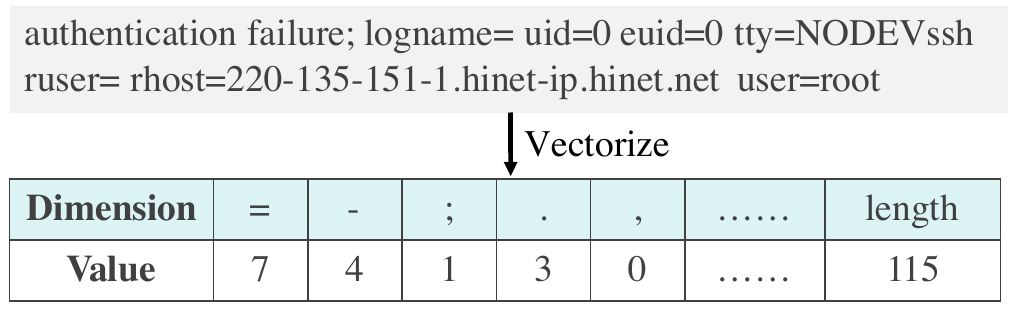}
\caption{Example of how to encode a log to a punctuation vector. As mentioned, \method{} encode this log to the vector [7, 4, 1, 3, 0, ……, 115].}
\label{fig6}
\vspace{-10pt}
\end{figure}

However, logs are hard to be embedded as there are too many unseen tokens in logs like tokens containing digits (e.g., "GIF534\_234"), Camel-Case words (e.g., "LogAnomalyDetection"), or some other special symbols (e.g., "\textbackslash{}etc\textbackslash{}wfs\textbackslash{}"), which cause out-of-vocabulary (OOV) problem easily. Based on investigation, we found that logs of the same template tend to have similar punctuation distribution, while logs of different templates tend to have different punctuation distribution. So we decide to utilize char-level embedding to encode logs into punctuation vectors. As shown in Fig \ref{fig6}, the vector has a dimension of m+1, where m denotes the number of selected punctuation features. The size of the first m dimensions of the vector reflects the number of occurrences of the corresponding punctuation in the log, while the last dimension of the vector represents the length of the log string. In \method{}, we select the 39 most commonly used punctuation as punctuation features.

Before calling Faiss, we use LLM to generate punctuation vectors, and by setting prompt statements, LLM determines the number $K$ of candidate templates for Faiss in advance according to massive background knowledge and generates a ranking of candidate templates according to the semantic similarity of vector contexts. After LLM encodes log $l_i$ into a punctuation vector $v_i$, the vectors are normalized to have a mean of 0 and variance of 1. This ensures that the importance of each dimension is measured on the same scale. Next, \method{} uses Faiss to recall the $k$-nearest-neighbor templates for log vector $v_i$. Only these templates need to be considered for future similarity calculations. 

\subsection{Step 4. Match}\label{Match}

Only one of the $k$-nearest-neighbor templates will be a candidate in the recall process. We chose Entropy-LCS, an information entropy-improved LCS to choose the most suitable template as the candidate. We calculate the longest common subsequence $\gamma$ in $k$-nearest-neighbor templates $T$, which may either be non-empty or empty. Subsequently, For each $t$ in $T$, we compare it with the LCS $\gamma$ and record the location of the token that is not the same. Then, we calculate the information entropy by $E(x) = -\sum_{i} p(x_i) \log p(x_i)$ for these position in $t'$. Following this, an exhaustive analysis is conducted over the positions of all divergent tokens, during which the information entropy and the list of tokens at these positions are calculated for all logs. The process of finding variables is the following: 
\begin{equation}
\text{V} = 
\begin{cases} 
\text{Yes}, & \text{if } -\sum_{t \in T} \frac{f(t)}{N} \log_2 \frac{f(t)}{N} > \theta \\
\text{No}, & \text{if } -\sum_{t \in T} \frac{f(t)}{N} \log_2 \frac{f(t)}{N} \leq \theta 
\end{cases}
\end{equation}
where $V$ donates variation point determination, The calculation of entropy $H$ is based on the distribution of tokens occurring at a specific position. $T$ is the set of tokens at a specific position. $f(t)$ is the frequency of token $t$ in all logs. $N$ is the total number of logs. $\theta$ is the threshold value. The decision for variables is $Yes$ if the calculated entropy exceeds the threshold $\theta$, otherwise $No$.

\subsection{Step 5. Update}\label{Update}
If the most suitable template is returned by section~\ref{Match}, \method{} will update the corresponding template in the template list. Tokens that are not present in the Entropy-LCS will be replaced by a special symbol (e.g., “$<$*$>$”) in the updated template. 
    
In case section~\ref{Match} does not return a suitable template, \method{} will consider the log as a new template and save it in the dynamic dictionary with the log abstract extracted from section~\ref{Retrieve} as an index. Additionally, the vector obtained from section~\ref{Recall} will be added to the Faiss index associated with this new template.

\begin{table*}[!t]
\caption{Summary of public datasets Loghub.}
\vspace{-5pt}
\label{summary_of_public_datasets}
\begin{center}
\resizebox{\textwidth}{!}{
\begin{tabular}{ccccccccc}
   \toprule
    &Dataset&Description&Time Span&Data Size&Logs&Template (total)&Template (2k) \\
    \midrule
    \multirow{5}{*}{\textcolor{red}{Distributed system} logs}
    &HDFS&Hadoop distributed file system log&38.7 hours&1.47 GB&11,175,629&30&14 \\
    &Hadoop&Hadoop mapreduce job log&N.A.&48.61 MB&394,308&298&114 \\
    &Spark&Spark job log&N.A.&2.75 GB&33,236,604&456&36 \\
    &ZooKeeper&ZooKeeper service log&26.7 days&9.95 MB&74,380&95&50 \\
    &OpenStack&OpenStack software log&N.A.&60.01 MB&207.820&51&43 \\
    \midrule

    \multirow{3}{*}{\textcolor{red}{Supercomputer} logs}
    &BGL&Blue Gene/L supercomputer log&214.7 days&708.7b MB&4,747,963&619&120 \\
    &HPC&High performance cluster log&N.A.&32.00 MB&433,489&104&46 \\
    &Thunderbird&Thunderbird supercomputer log&244 days&29.60 GB&211,212,192&4,040&149 \\
    \midrule

    \multirow{3}{*}{\textcolor{red}{Operating system} logs}
    &Windows&Windows event log&226.7 days&26.09 GB&114,608,388&4,833&50 \\
    &Linux&Linux system log&263.9 days&2.25 MB&25,567&488&118 \\
    &Mac&Mac OS log&7.0 days&16.09 MB&117.283&2,214&341 \\
    \midrule
   
    \multirow{2}{*}{\textcolor{red}{Mobile system} logs}
    &Android&Android framework log&N.A.&3.38 GB&30,348,042&76,923&166 \\
    &HealthApp&Health App log&10.5 days&22.44 MB&253,395&220&75 \\
    \midrule
   
    \multirow{2}{*}{\textcolor{red}{Server application} logs}
    &Apache&Apache server error log&263.9 days&4.90 MB&56,481&44&6 \\
    &OpenSSH&OpenSSH server log&28.4 days&70.02 MB&655,146&62&27 \\
    \midrule
   
    \multirow{1}{*}{\textcolor{red}{Standalone software} logs}
    &Proxifier&Proxifier software log&N.A.&2.42 MB&21,329&9&8 \\
    
   \bottomrule
\end{tabular}
}
\end{center}
\end{table*}

\section{Experiment}

\subsection{Research Questions}
To verify the effectiveness and efficiency of \method{}, we conduct extensive experiments on both loghub and \bench{}, aiming to answer the following questions:

\textbf{RQ1:} How effective is \method{} on public Loghub and our \bench{}?

\textbf{RQ2:} How efficient Is \method{} with the change of log and template volume?

\textbf{RQ3:} How does \method{} perform without LLM and Faiss to retrieve and recall by constructing the special dynamic dictionary?

\textbf{RQ4:} How do we set suitable hyperparameters in recall and E-LCS process affect the performance of \method{}?

\subsection{Implementation and Environment}
In our experiments, we set up a Virtual Machine (VM) with 64 Intel Core I5 CPU @ 2.0GHz processors and 16GB RAM. The operating system is Ubuntu-20.04. For strong keyword representation, cross-linguistic understanding, and
cross-semantic comprehension of LLM, we choose ChatGPT-4 as the base of \method{}. There are three hyperparameters provided for adjustment in \bench{}: the number of the nearest neighbor templates $k$ is set to $5$, the recall threshold $\tau$ is set to $0.5$, and the matching threshold $\theta$ is set to $4.5$. These default values remain consistent in most experiments. Using these default parameters, \method{} achieves high accuracy and reliability across various datasets.




\subsection{Datasets}\label{Dataset}
\method{} is evaluated on the public LogHub and the industrial dataset \bench{} collected by us. LogHub contains 16 public datasets. The data from LogHub are collected from common open-source components, such as Spark, Zookeeper, and Apache. For each log source, 2000 logs are sampled and labeled manually. As for \bench{}, the data was collected from actual business scenarios in the fields of finance, communication, and manufacturing, spanning over 30 days, with over 300000 logs per domain, ranging from mixed logs across languages to single language logs. We have presented a summary of our industrial datasets \bench{} and public Loghub datasets in Table~\ref{summary_of_industrial_datasets} and Table~\ref{summary_of_public_datasets}, respectively.

\begin{table}[htbp]
\caption{Summary of industrial datasets \bench{}.}
\label{summary_of_industrial_datasets}
\begin{center}
\resizebox{\linewidth}{!}{
\begin{tabular}{ccccc}
   \toprule
    Dataset&Templates&Avg. Len.&\makecell{Various Lengths \\ Proportion}&Logs\\
   \midrule
    Finance&\textcolor{red}{438}&75.21&40.8\%&303579\\
    Communication&\textcolor{red}{192}&47.36&60.9\%&414347\\
    Manufacturing&\textcolor{red}{70}&41.98&4.74\%&306218\\
   \bottomrule
\end{tabular}
}
\end{center}
\end{table}

\subsection{Baselines and Metrics}

As for baselines, We choose Drain\cite{drain}, IPLOM\cite{iplom}, and Spell\cite{spell} as our baselines. They work well on the LogHub. To quantify the effectiveness of \method{}, we leverage F-measure and Group Accuracy (GA) consistent with prior studies\cite{drain, iplom, spell}. F-measure is a template-level metric that focuses on the ratio of correctly grouped templates and GA is computed as the ratio of correctly grouped log messages to the total count of log messages. Specifically, We also measure the execution time in seconds and compare \method{} with other parsers in terms of efficiency.

\subsection{\textbf{RQ1:} How effective is \method{} on public loghub and our \bench{}?}

\begin{table*}[htbp]
\caption{F-measure and GA on public logHub.}
\vspace{-5pt}
\label{public_data_accuracy}

\begin{center}
\resizebox{0.75\textwidth}{!}{
 \small
\begin{tabular}{ccccccc>{\columncolor{gray!25}}c>{\columncolor{gray!25}}c}
   \toprule
   
   \multirow{2}*{Dataset}&\multicolumn{2}{c}{Drain}&\multicolumn{2}{c}{Spell}&\multicolumn{2}{c}{IPLOM}&\multicolumn{2}{c}{\method{}}\\
   
   \cmidrule(lr){2-3}\cmidrule(lr){4-5}\cmidrule(lr){6-7}\cmidrule(lr){8-9}
   
   &F-measure&PA&F-measure&PA&F-measure&PA&F-measure&PA\\
    
   \midrule
    HDFS&0.999&0.998&\textbf{1}&\textbf{1}&\textbf{1}&\textbf{1}&0.999&0.998\\
    Hadoop&\textbf{0.999}&0.948&0.920&0.778&0.996&0.954&\textbf{0.999}&\textbf{0.987}\\
    Spark&0.992&0.920&0.991&0.905&0.992&0.920&\textbf{0.999}&\textbf{0.997}\\
    Zookeeper&\textbf{0.999}&0.967&\textbf{0.999}&0.964&\textbf{0.999}&0.962&\textbf{0.999}&\textbf{0.991}\\
    BGL&\textbf{0.999}&\textbf{0.963}&0.957&0.787&\textbf{0.999}&0.939&\textbf{0.999}&0.949\\
    HPC&0.991&0.887&0.986&0.654&0.978&0.829&\textbf{0.992}&\textbf{0.903}\\
    Thunderbird&\textbf{0.999}&0.955&\textbf{0.994}&0.844&\textbf{0.999}&0.663&\textbf{0.999}&0.946\\
    Windows&\textbf{0.999}&\textbf{0.997}&\textbf{0.999}&0.683&0.995&0.568&\textbf{0.999}&0.994\\
    Linux&\textbf{0.992}&0.690&0.937&0.605&0.964&0.672&\textbf{0.994}&\textbf{0.868}\\
    Andriod&\textbf{0.996}&0.911&0.992&\textbf{0.919}&0.949&0.712&0.991&0.801\\
    HealthApp&0.918&0.780&0.887&0.639&0.958&\textbf{0.822}&\textbf{0.971}&0.819\\
    Apache&\textbf{1}&\textbf{1}&\textbf{1}&\textbf{1}&\textbf{1}&\textbf{1}&\textbf{1}&\textbf{1}\\
    Proxifier&0.785&0.527&0.832&0.527&0.786&0.517&\textbf{0.999}&\textbf{0.977}\\
    OpenSSH&\textbf{0.999}&0.788&0.918&0.554&0.998&0.540&\textbf{0.999}&\textbf{0.814}\\
    OpenStack&0.993&0.733&0.994&0.764&0.909&0.331&\textbf{1}&\textbf{1}\\
    Mac&0.975&0.787&0.963&0.757&0.957&0.671&\textbf{0.977}&\textbf{0.850}\\
    
    \midrule
    \textbf{Average}&0.977&0.866&0.961&0.774&0.968&0.756&\textbf{0.995}&\textbf{0.931}\\
   \bottomrule
\end{tabular}
}
\end{center}
\end{table*}

\begin{table*}[htbp]
\caption{F-measure and GA on \bench{}.}
\begin{center}
\resizebox{0.75\textwidth}{!}{
\begin{tabular}{ccccccc>{\columncolor{gray!25}}c>{\columncolor{gray!25}}c}
   \toprule
   \multirow{2}*{Dataset}&\multicolumn{2}{c}{Drain}&\multicolumn{2}{c}{Spell}&\multicolumn{2}{c}{IPLOM}&\multicolumn{2}{c}{\method{}}\\
   \cmidrule(lr){2-3}\cmidrule(lr){4-5}\cmidrule(lr){6-7}\cmidrule(lr){8-9}
   &F-measure&PA&F-measure&PA&F-measure&PA&F-measure&PA\\
   \midrule
    Finance&0.469&0.380&0.090&0.249&0.469&0.362&\textbf{0.988}&\textbf{0.626}\\
    Communication&0.863&0.268&0.877&0.171&0.898&0.273&\textbf{0.920}&\textbf{0.423}\\
    Manufacturing&0.984&0.793&0.906&0.193&0.569&0.469&\textbf{0.998}&\textbf{0.832}\\
    \midrule   \textbf{Average}&0.772&0.481&0.090&0.249&0.645&0.368&\textbf{0.969}&\textbf{0.627}\\
   \bottomrule
\end{tabular}
}
\vspace{-10pt}
\end{center}
\end{table*}

\subsubsection{Public loghub results}
Table~\ref{public_data_accuracy} demonstrates the effectiveness of \method{} and three baselines. The best results for each dataset have been highlighted. The table shows that though baselines have already achieved relatively good results in public loghub, \method{} still outperforms the other algorithms in terms of F-measure on 14 out of 16 datasets and GA on 10 out of 16 datasets. Overall, \method{} achieves the highest average scores for both F-measure and GA, with a 1.8\% improvement in F-measure and a 6.5\% improvement in GA. One of the most notable achievements of \method{} is its performance on the Proxifier dataset. This dataset is particularly suitable for \method{}, as it contains 47.8\% similar logs with semantic differences and 47.35\% logs with varying lengths. 

\subsubsection{\bench{} results}
In \bench{}, the improvement of \method{} is more obvious. It achieved an average F-measure of 0.969, which is 25.5\% higher than the best result in the baselines, and an average PA of 0.627, which is 30.3\% higher than the best result in the baselines. The greatest effect improvement of \method{} was observed on the Finance dataset, which contains the maximum proportion of logs with different semantics and various lengths. Specifically, the proportions of such logs were 17.9\% and 40.8\%, respectively.

\subsection{\textbf{RQ2:} How efficient is \method{} with the change of log and template volume?}

To measure the efficiency of our proposed approach, we compare the execution time of \method{} with the baselines as the number of logs and templates varies. Since LogHub does not have a large scale of templates, the data for the experiment are randomly sampled from the LogHub and \bench{}. For fairness, we utilize only a single executor and run each log parser five times for the average execution time.

\begin{figure}[htp]
\centering
\includegraphics[width=0.9\linewidth]{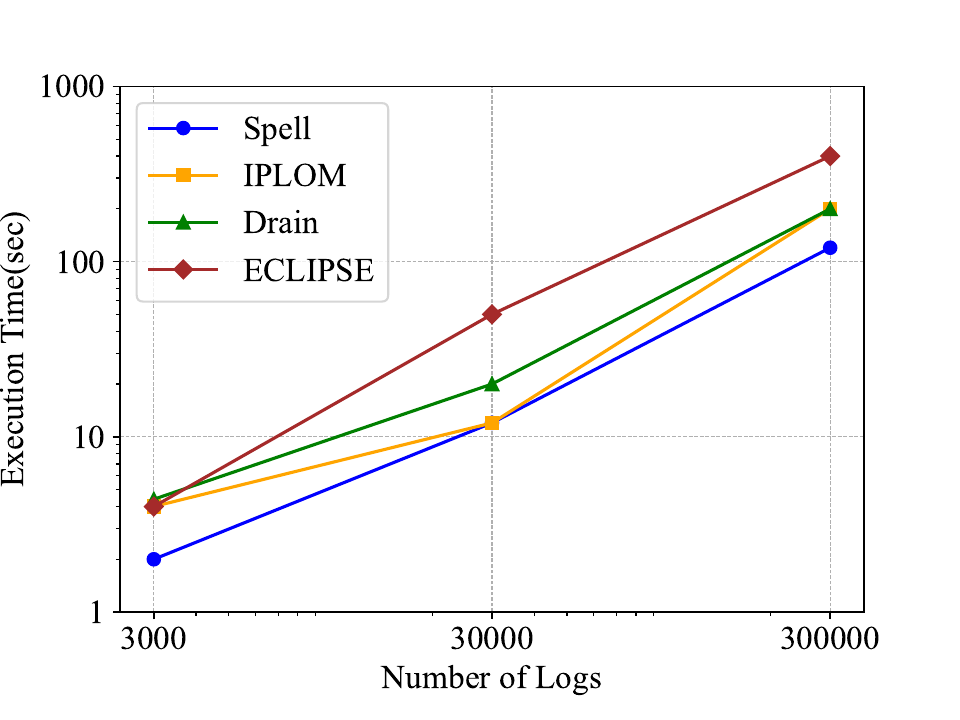}
\caption{Execution time comparison with baselines at different numbers of logs.}
\vspace{-10pt}
\label{fig:logtime}
\end{figure}

\begin{figure}[htp]
\centering
\includegraphics[width=0.9\linewidth]{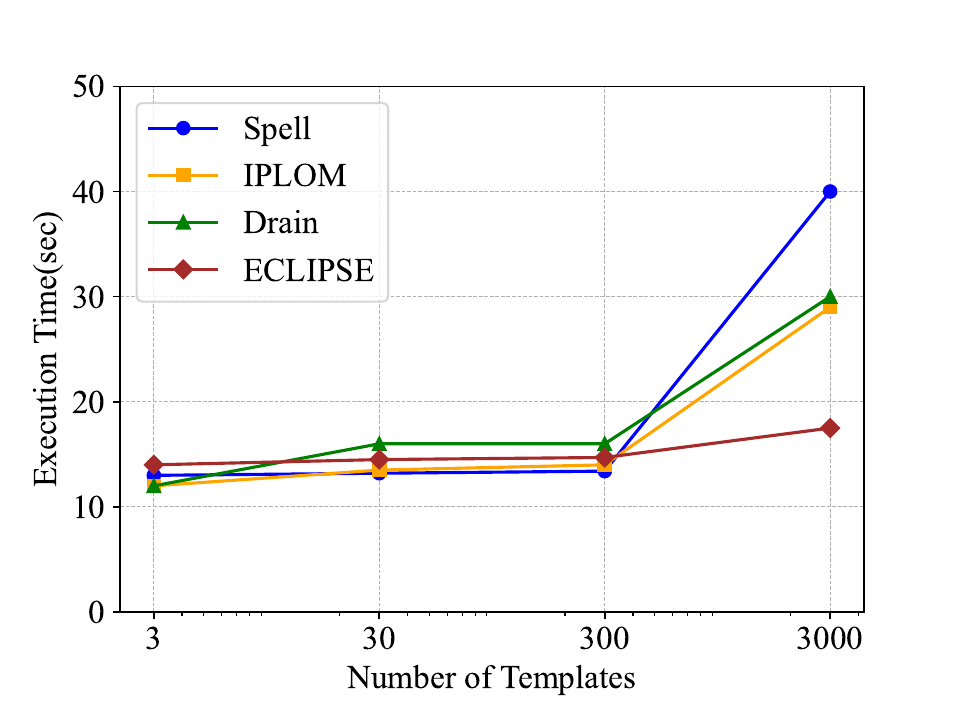}
\caption{Execution time comparison with baselines at different numbers of log templates.}
\label{fig:templatetime}
\vspace{-10pt}
\end{figure}

For the execution time comparison on the number of logs, we measure the execution time of each parser under the different scales $\{3000, 30000, 300000\}$ of log data in Figure~\ref{fig:templatetime}. Results demonstrate that \method{} achieves the lowest time consumption regardless of the log volume. Besides, as logs grow, \method{} becomes more efficient compared to baselines. Especially when the number of logs reaches $300000$, \method{} takes almost half the execution time of Spell. In Figure~\ref{fig:logtime}, the number of logs remains unchanged $(100000)$. Thus we conduct experiments under the different numbers $\{3, 30, 300, 3000\}$ of log templates. When the number of templates is small, \method{} is not as efficient as baselines. However, as the number of templates rises to a specific order of magnitude, \method{} shows a significant advantage in efficiency. Both experiments demonstrate that \method{} is far more efficient in industrial scenarios, especially when more log templates and larger log volumes.

\begin{figure}[htp]
\centering
\includegraphics[width=0.85\linewidth]{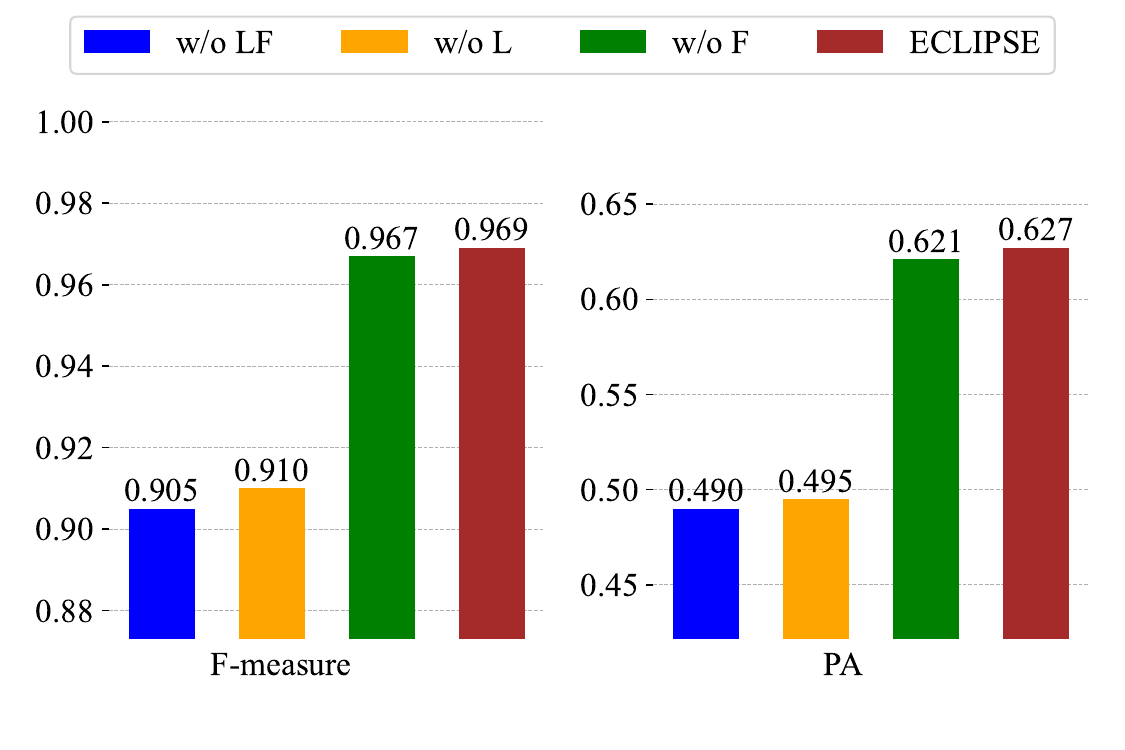}
\caption{Ablation study results on effectiveness.}
\label{fig:ablation_effect}
\end{figure}

\begin{figure}[htp]
\centering
\includegraphics[width=0.9\linewidth]{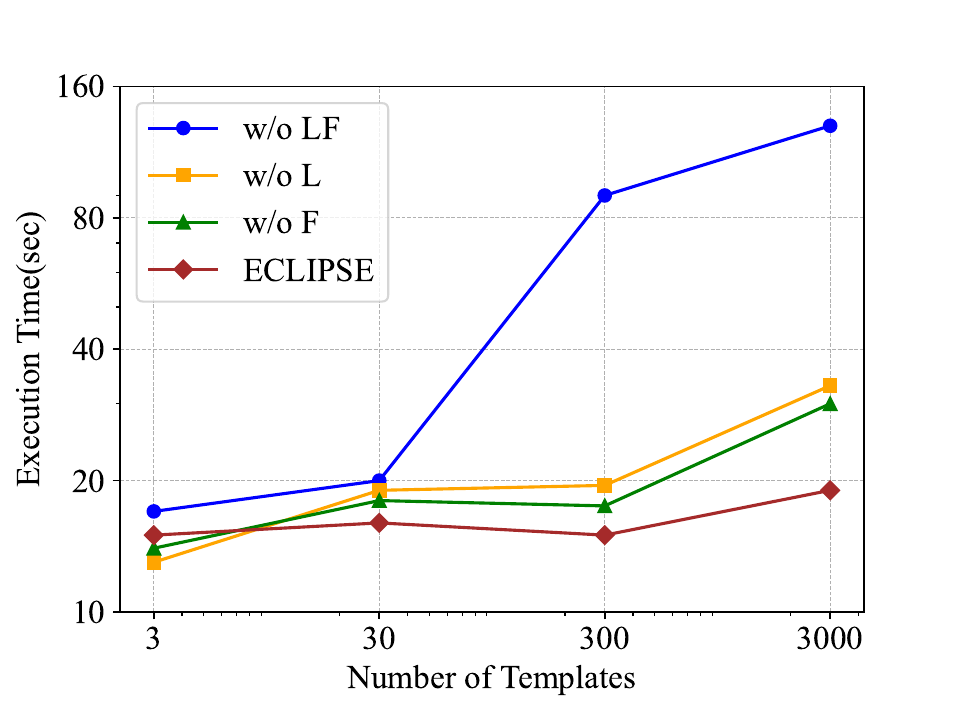}
\caption{Ablation study results on efficiency.}
\label{fig:ablation_efficiency}
\end{figure}

\subsection{\textbf{RQ3:} How does \method{} perform without LLM and Faiss to retrieve and recall by constructing the special dynamic dictionary?}


In this section, we conduct an ablation experiment to investigate the effectiveness and efficiency of LLM and Faiss in \method{}. To verify this, we compare \method{} with three variants of \method{}: 1) \method{} w/o LF (\method{} without LLM and Faiss); 2) \method{} w/o L (\method{} without LLM); 3) \method{} w/o F (\method{} without Faiss).

\subsubsection{Effect on Effectiveness}
We conduct the ablation experiment to verify their effects on effectiveness based on our industrial datasets \bench{} as it's closer to the real parsing situation. The results are shown in Figure~\ref{fig:ablation_effect}, where the label "$w/o$" means "without", "$kl$" means "keyword library" and "$fi$" means "Faiss". As we can see, the keyword library has a certain impact on the effectiveness, with about 6.7\% improvement on F-measure and 24.4\% improvement on PA, while Faiss hardly affects the parsing effectiveness, just as we expect. This proves that recall based on the Faiss has a very high recall rate.

\subsubsection{Effect on Efficency}
The efficiency of these two components varies with the number of templates. As shown in Figure~\ref{fig:ablation_efficiency}, without the keyword library and Faiss, the execution time of \method{} significantly increases with the increase in the number of templates. And with the keyword library and Faiss, their growth rate has significantly decreased. When the number of templates is only 3, there are not many differences between \method{} Base and other variants. But when the number of templates comes to 3000, \method{} Base can be 6.76 times faster than variant 1), 1.68 times faster than variant 2), and 1.47 times faster than variant 3), which shows that both keyword dictionary and Faiss can effectively improve the efficiency of \method{}.


   

   
    

    

\subsection{\textbf{RQ4:} How do we set suitable hyperparameters in recall and E-LCS process affect the performance of \method{}?}

In this section, we discuss the effect of three parameters: the number of the nearest neighbor templates $k$, the recall threshold $\tau$, and the matching threshold $\theta$. All experiments are conducted on Manufacturing, which is the subset of \bench{}.

For the validation of parameter $k$, we set three variables $\{5,10,15\}$ and keep the other parameters constant. The results in Table~\ref{parameter_k_ablation} show that the performance is less affected by the changes of $k$. As the parameter values become larger, the performance decreases a little, which demonstrates the robustness of the model.

For the validation of $\tau$, we set three variables $\{0.1,0.5,0.9\}$ and keep the other parameters constant. The results prove that the model is more sensitive to $\tau$. As the value of $\tau$ becomes larger, the performance is first superior and then inferior. Our optimal experimental setting is 0.5 for $\tau$. For $\theta$, we set five variables $\{3.5, 4.0, 4.5, 5.0,\\ 5.5\}$ and keep the other parameters constant. The results show that the model is less influenced by the $\theta$. As the $\theta$ becomes larger, the performance increases and then decreases. Our best setting is 0.6 for $\theta$.

\begin{table}[tbp]
\caption{Performance under different $k$.}
\label{parameter_k_ablation}
\vspace{-5pt}
\begin{center}
\scalebox{1}{
\begin{tabular}{cccccc}
   \toprule
    Dataset&$k$&$\tau$&$\theta$&F-measure&GA\\
   \midrule
   \quad&5&0.5&4.5&\textbf{0.998} &\textbf{0.832}\\
    Manufacturing &10&0.5&4.5&0.988&0.821\\
   \quad&15&0.5&4.5&0.986&0.817\\
   \bottomrule
\end{tabular}
}
\end{center}
\vspace{-5pt}
\end{table}

\begin{table}[htbp]
\caption{Performance under different $\tau$.}
\vspace{-5pt}
\label{parameter_tv_ablation}
\begin{center}
\scalebox{1}{
\begin{tabular}{cccccc}
   \toprule
    Dataset&$k$&$\tau$&$\theta$&F-measure&GA\\
   \midrule
   \quad&5  &0.1&4.5 &0.957 &0.399\\
    Manufacturing &5&0.5 &4.5 &\textbf{0.998} &\textbf{0.832}\\
   \quad&5  &0.9  &4.5  &0.987  &0.798\\
   \bottomrule
\end{tabular}
}
\end{center}
\vspace{-5pt}
\end{table}

\begin{table}[htbp]
\caption{Performance under different $\theta$.}
\label{parameter_ts_ablation}
\begin{center}
\scalebox{1}{
\begin{tabular}{cccccc}
   \toprule
    Dataset&$k$&$\tau$&$\theta$&F-measure&GA\\
   \midrule
   \quad &5& 0.5& 3.5&0.985 & 0.761\\
    \quad &5&0.5&4.0 & 0.988 & 0.787\\
   Manufacturing &5&0.5 &4.5&\textbf{0.998}&\textbf{0.832}\\
   \quad&5&0.5&5.0 &0.996 & 0.826\\
   \quad&5&0.5&5.5 &0.979 & 0.803\\
   \bottomrule
\end{tabular}
}
\end{center}
\end{table}

\section{THREATS TO VALIDITY}
The following major threats to validity are identified: \textbf{1) Under-partitioning risk}. \method{} may be at risk of under-partitioning, which means merging different templates into one. However, we have found LLM and Faiss to construct a dynamic dictionary help a lot in reducing this risk, and experiments on LogHub and \bench{} show \method{} performs well, and this problem rarely occurs in real. \textbf{2) Insufficient attention to infrequent templates.} It is an easily overlooked problem in existing work, as the impact of frequent templates will have a greater weight than that of infrequent templates in evaluation metrics. However, logging infrequent templates may be more important in real industrial scenarios because serious events often occur infrequently. So this will be a key research direction for our future work.


\section{Conclusion}
In this paper, we propose a robust log parsing method called \method{}. This method is designed to parse logs in a streaming manner, providing both accuracy and efficiency even in the most complex scenarios. When parsing a new log, \method{} first quickly recalls a specified number of templates based on a dynamic library driven by LLM and Faiss, which has been proven effective through ablation experiments. Then, \method{} utilizes our proposed Entropy-LCS to match the most suitable template and update it in real time. We evaluate \method{} on both public LogHub and our \bench{}. The results show that while baselines perform well only on public datasets, \method{} not only outperforms them on public datasets but also excels on industrial datasets. In terms of efficiency, \method{} also has clear advantages, especially when the number of templates reaches a significantly high level, allowing it to parse logs quickly and accurately.

\clearpage
\bibliographystyle{ACM-Reference-Format}
\bibliography{ref}

\end{document}